# On the Aggregation State of Synergistic Antimicrobial Peptides

Jacob M. Remington,[a] Chenyi Liao,[a] Mona Sharafi,[a] Emma Ste. Marie,[a,b] Jonathon B. Ferrell,[a] Robert Hondal,[a,b] Matthew J. Wargo,[c] Severin T. Schneebeli,[a] and Jianing Li[a]*

a.  Department of Chemistry, University of Vermont, Burlington, VT 05405.

b.  Department of Biochemistry, University of Vermont, Burlington, VT 05405.

c.  Department of Microbiology and Molecular Genetics, University of Vermont, Burlington, VT 05405.

**Corresponding Author**

*Jianing Li (jianing.li@uvm.edu)

**ABSTRACT.** By integrating various simulation and experimental techniques, we discovered that antimicrobial peptides (AMPs) may achieve synergy at an optimal concentration and ratio, which can be caused by aggregation of the synergistic peptides. On multiple time and length scales, our studies obtain novel evidence of how peptide co-aggregation in solution can affect disruption of membranes by synergistic AMPs. Our findings provide crucial details about the complex molecular origins of AMP synergy, which will help guide the future development of synergistic AMPs as well as applications of anti-infective peptide cocktail therapies.

Antibiotic resistance is one of the biggest threats to global health and food production. It leads to dangerous infections, long hospital stays, high medical costs, and significantly increased mortality.[1,2] Additional to solutions like conventional antibiotics and vaccines, antimicrobial peptides (AMPs), especially broad-spectrum antibacterial ones, have emerged to aid our battle against challenging bacterial infections and multi-drug resistance. Over 12,000 antimicrobial AMPs have been identified from natural sources or via bioorganic synthesis.[3–5] Despite great variation in the sequence, length, and structure, many AMPs have their antimicrobial activity largely attributed to the ability to induce membrane disruption or poration.[6,7] Several mechanisms of action have been proposed, based on experimental evidence,[8,9] as well as molecular dynamics (MD) simulations[10–12] and accelerated techniques[13–16] using either all-atom or coarse-grained models.[17]

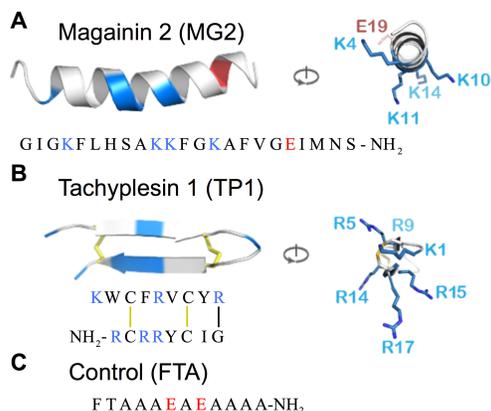

**Figure 1.** The sequences and structures of MG2 (**A**), TP1 (**B**) (PDB IDs: 2MAG and 2RTV), and the sequence of a non-AMP called FTA (**C**). Positively charged residues are colored blue while negatively charged ones are red. The 23-residue MG2 is helical with a +4 charge, and the 17-residue TP1 with a +7 charge forms an antiparallel β-sheet with two S-S linkages. Non-AMP FTA was chosen as a control system given the similar length to TP1.

While earlier studies almost exclusively focused on individual peptides, most natural AMPs exist in mixtures.[18,19] It has been long hypothesized that combinations of two or more types of AMPs may generate synergism,[20,21] where the effect of combined AMPs is stronger than that of each component in the equivalent dose. However, only few examples of synergistic AMPs have been identified,[22] mainly due to limited knowledge about the synergy mechanisms. In this work, we obtained novel evidence from simulations and experiments to show how peptide aggregation in solution can affect disruption of membranes by synergistic AMPs, with a focus on two prototypical AMPs — Magainin 2 (MG2) and Tachyplesin 1 (TP1) — which are similar in the molecular mass but distinct in sequence and structure (Figure 1). The drastic structural differences between these two peptides contrasts MG2 and TP1 synergy from recent [23,24] and past work[25–27] on MG2 and PGLa synergy wherein both peptides are helical and found to self-assemble on the membrane surface to form transmembrane pores. Our findings are significant in advancing current knowledge of AMP synergy mechanisms between peptides with different secondary structure elements, and in providing a new route for seeking synergistic AMPs.

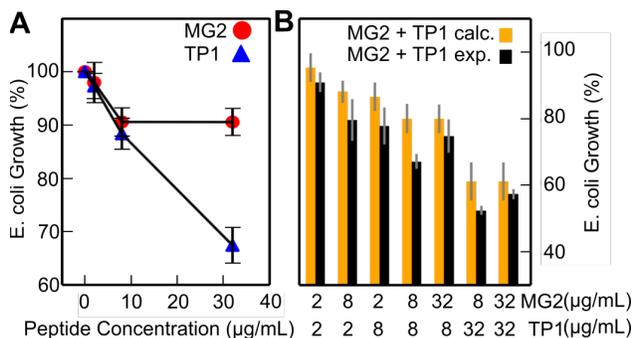

**Figure 2.** *E. coli* growth assay to test MG2-TP1 synergy. (**A**) Plots of pure MG2 or TP1 concentration for *E. coli* inhibition. (**B**) Activity of MG2-TP1 combinations (black), compared with the calculated E. coli growth from independent application of MG2 and TP1 assuming no synergy (orange). Error bars (1σ) were determined from three independent trials.

MG2 and TP1 have been long known to kill bacteria via poration of bacterial membranes.[28–32] Their synergy was previously suggested at 1.25 μM MG2 and 0.1 μM TP1, the minimum concentrations to inhibit *E. coli* growth.[33] Herein, we performed a systematic investigation, by directly measuring the percentages of *E. coli* growth inhibition at different concentrations and ratios. The quantitative nature of our results allowed us, for the first time, to investigate and optimize the actual ratios of the synergistic peptides, which has not been accomplished yet with the coarser measurements of minimum inhibition concentration. We discovered that maximum synergy was achieved with a 1:1 ratio of MG2 and TP1, while MG2 was present in large excess in the prior study.[33]

We started with *E. coli* growth assays, which showed that both peptides inhibited bacterial growth when present in concentrations as low as 2 µg/mL (<1 µM). Furthermore, we found that *E. coli* growth decayed with increasing peptide concentration (Figure 2A), with TP1 appearing more potent than MG2. Synergy was then quantified by comparing experimentally measured *E. coli* growth (black bars in Figure 2B, with doses of 1:1, 1:4, and 4:1 MG2: TP1 mass ratio, raw data in Table S1-S3) to the expected E. coli growth ($G_{\text{calculated}}$) assuming the peptides act independently to inhibit bacterial growth (Eq. 1, orange bars in Figure 2B).

$$G_{\text{calculated}}([\text{MG2}], [\text{TP1}]) = G([\text{MG2}])G([\text{TP1}]) \qquad \text{Eq. 1}$$

In Eq. 1 $G([\text{MG2}])$ and $G([\text{TP1}])$ are the measured concentration dependent *E. coli* growth of the individual peptides. While synergy was observed in nearly all of pairs of concentrations across all trials, (Tables S1-S3) only the 1:1 ratio with 8 µg/mL of each peptide showed a statistically significant difference at a 95% confidence level (see Supporting Information). This combination had a 12.9 ± 3.9% higher activity than expected without synergy. However, excess of MG2 did not appear to enhance synergy, as becomes clear when one, for example, compares the *E. coli* growth at 32 µg/mL MG2 and 8 µg/mL TP1 versus at 8 µg/mL MG2 and 32 µg/mL TP1 (Figure 2B). This later finding contrasts findings from previous work that found higher MIC at higher ratios of MG2 to TP1.[33] Generally, these results reveal a dependence of MG2-TP1 synergy on peptide concentrations and ratios.

With the synergy confirmed, we next discovered that MG2 and TP1 co-aggregate in solution, using Dynamic Light Scattering (DLS) spectra and large-scale, all-atom MD simulations. At the tested concentrations (2.0 × $10^{-8}$ to $10^{-11}$ M), pure MG2 or TP1 aggregated into particles of hydrodynamic diameters $R_h$ > 80 nm (Figures 3D and S1). However, our DLS data indicate that the formation of large peptide aggregates is generally reduced when the MG2 and TP1 peptides are mixed in a 1:1 ratio with the $R_h$ of the particles formed in the range of 13 - 200 nm (Figure 3D), which differ from the aggregates formed by either MG2 or TP1 alone. It is noted that the aggregates larger than 200 nm in Figure 3D are only present in very low numbers (since the DLS intensity depends on the 6th power of the particle size) and are thus unlikely to be the major active antimicrobial species in solution. These data suggest that within the sub-micromolar concentration range, MG2 and TP1 co-aggregate into smaller particles than the ones formed by individual peptides.

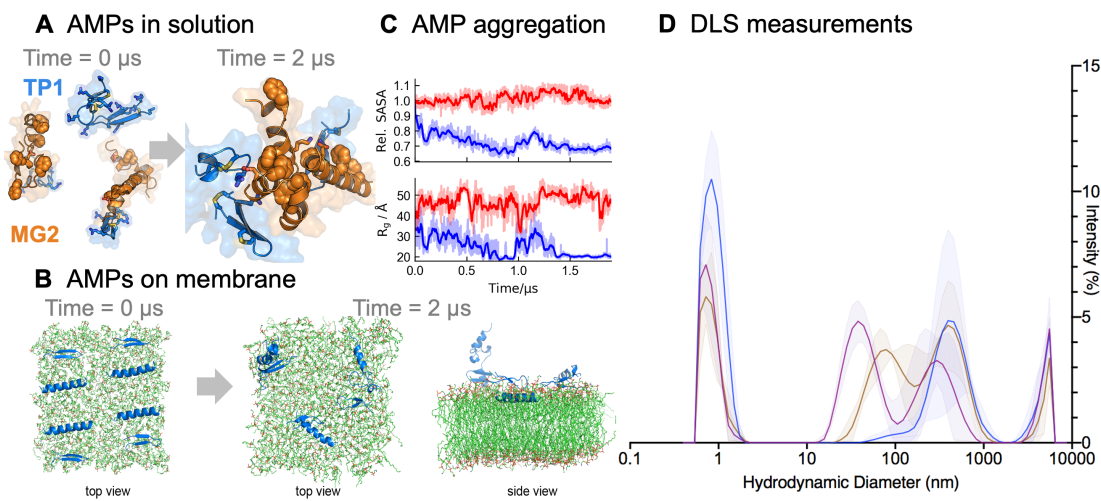

**Fig 3. (A-B)** Comparison of MG2 and TP1 co-aggregation in the solution and membrane environments. In panel B the initial frame for the AMP's associating on the membrane is shown. **(C)** The relative SASA and radius of gyration ($R_g$) for MG2 and TP1 co-aggregation in the solution (blue) and on the membrane (red). **(D)** DLS spectra of TP1 (blue), MG2 (brown), and a 1:1 TP1/MG2 mixture (magenta). The total AMP concentration ([AMP]$_{total}$) was 2.0 nM in each case. Error bands (1σ) are from three DLS measurements. The DLS results confirm that smaller aggregates are formed with the mixture of the peptides, compared to when the pure MG2 or TP1 is used with a comparable concentration.

Consistent with the DLS evidence, our MD simulations show that the MG2 and TP1 peptides co-aggregate in solution on the microsecond timescale. We simulated four MG2 (either helical or coiled) and four TP1 (either cyclic or linear) peptides arbitrarily placed in the solution), and observed TP1 and MG2 forming hetero-oligomers of various sizes in all of our simulations (Figure S2 and S3). While MD simulations and DLS spectra agreed on the higher aggregation tendency of MG2 (compared to TP1) in pure peptide solutions, the helical MG2 and cyclic TP1 formed the most ordered and stable aggregates — with MG2 in the core and TP1 in the vicinity (Figure 3A). Loose complexes formed in the early stage with a radius of gyration, $R_g$ = 38 Å, and then gradually turned into a stable hetero-oligomer (Figure 3A) toward the end of the simulations, with $R_g$ reducing to as low as 18-25 Å and the solvent-accessible surface area (SASA) decreasing by 30 to 35% within 2 μs. Further simulation analysis provided more structural insight: like many amphipathic AMPs,[34] MG2 (Figure 1A) used one hydrophobic surface (F5, F12, and F16) for aggregation and the other surface (K4, K10, K11, K14, and E19) for solvent exposure; due to the alternating hydrophobic and cationic residues (Figure 1B), TP1 was inclined to bind E19 of MG2 and attach to the MG2 core in a nonspecific fashion (Figure 3A).

With MG2-TP1 co-aggregation confirmed both experimentally and computationally, we simulated the detailed mechanism of membrane poration and provided evidence that the MG2-TP1 hetero-oligomers formed in the solution may damage bacterial membranes as an entity. AMP-induced pore formation — defined here as water channels that span the membrane — can be a central event of AMP mechanisms. Although it has been speculated that AMPs aggregate on the bacterial membrane surface to form membrane pores, MG2 and TP1 co-aggregation on the membrane surface likely occurs on a longer timescale than the corresponding co-aggregation in solution. We simulated four helical MG2 and four cyclic TP1 peptides dispersed on the surface of lipid bilayers (PE:PG 3:1, mass percent), but no consistent aggregation was observed on the membrane surface during the 2-μs simulations (Figure 3B), as indicated by the large $R_g$ and SASA values of the peptides (Figure 3C). This finding contrasts the much faster MG2-TP1 hetero-oligomer formation in solution (Figure 3A), which supports the possibility that MG2 and TP1 act on the bacterial membrane in the hetero-oligomeric form.

To gain further evidence, we utilized an enhanced sampling protocol to (1) surmount the energy barrier associated with the MG2-TP1 oligomer embedding in the membrane and (2) sample relevant complex conformations in membrane-disrupted states. Indeed, the process of membrane poration often occurs on larger timescales than what has been simulated.[35,36] It is even harder for complex systems with multiple peptide and lipid species in the bacterial membrane models. To obtain relevant conformations of peptide aggregates in a membrane-disrupted state, we have designed a protocol that employed alternating steered and unbiased MD simulations to mimic the process of peptides colliding into the elastic lipid bilayer. Mechanistically, this approach was used to reveal how the experimentally observed hetero-oligomers may interact selectively with the bacterial membrane to form a pore. In each step, we pulled the hetero-oligomer along the negative Z-axis direction at a constant velocity applied to the peptide backbones for 40 ps (Figure S4), followed by an 8-ns relaxation of the peptide-lipid interactions. After ~20 steps a pore emerged in the membrane (indicated by water penetration), an unbiased MD simulation was run to sample stable MG2-TP1 conformations in the membrane pore (Figure 4). Two membrane models to mimic different bacterial membranes were tested (mass percent): (1) PE:PG 3:1 and (2) PE:PG:Cardiolipin 65:10:25, which displayed stable membrane pores under the simulation protocol. These pores remained during the final stage for over 100 ns and were observed in replica simulations (Figure S5), suggesting converged simulations with limited fluctuations of stable pore structures. Using similar protocols, however, we did not see pore formation in the simulations of MG2-TP1 oligomer in a mammalian membrane model (POPC:cholesterol 9:1, Figure S6) and a non-AMP peptide, FTA (Figure 1C) oligomer in the membrane (Figure 4A1). Both of these two results suggest that the mechanical forces alone (Figure S4) were insufficient to disrupt the membrane integrity. Hence, these results highlight the ability of our simulation protocol to differentiate between peptide oligomers which disrupt bacterial membranes and those that do not, providing evidence that the sampled conformations are biophysically relevant.

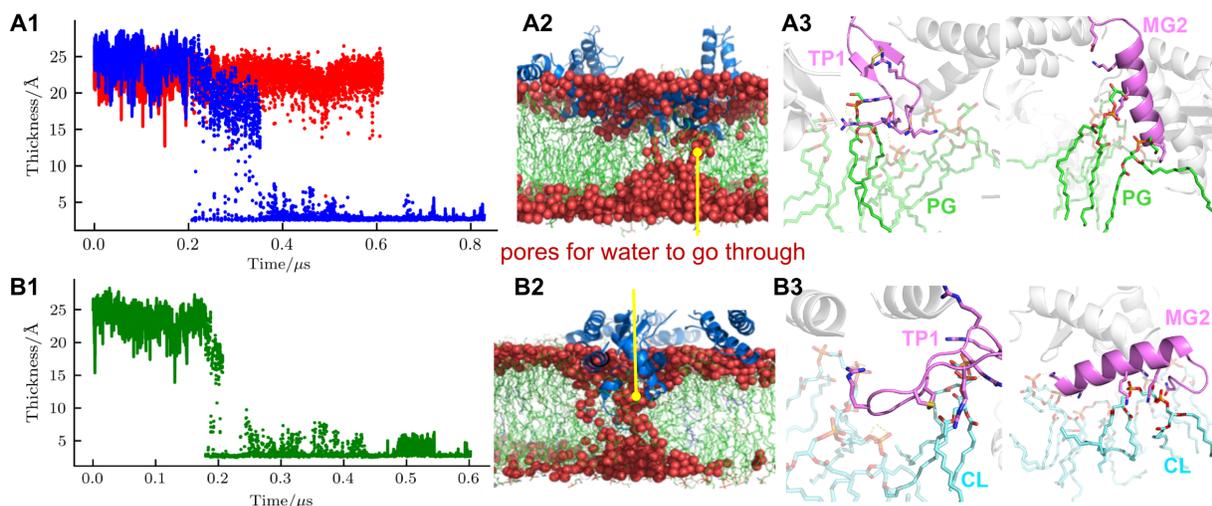

**Fig 4.** (A) Membrane thickness as judged by the minimal distance between waters above and below the POPE:POPG 3:1 ratio membrane during equilibrium (solid line) and SMD (dots) runs. The MG2-TP1 hetero-tetramer (blue) forms a stable pore at 350 ns while the control sequence (red) doesn't form a pore within the 600 ns. The final snapshot illustrating the membrane pore formed by MG2-TP1. The orientation of a MG2 monomer at the membrane/water interface exposes hydrophobic residues to the membrane tails and charged/polar ones to the water and membrane heads. (B) Membrane thickness for the MG2-TP1 hetero-tetramer in a cardiolipin- rich membrane model. Final snapshots of SMD and equilibrium MD runs for pore formation in a cardiolipin rich bacterial membrane model. Side chain interactions of TP1 with cardiolipin head groups.

Further examination of the hetero-oligomer conformations and contacts with the lipids suggest that synergy between MG2 and TP1 may arise from forming active hetero-oligomers. (1) The hetero-oligomers provide the specific molecular scaffold for membrane disruption, similar to what was hypothesized about amyloid pores.[37–40] Specifically, the MG2-TP1 hetero-oligomers remained compact during the simulated process of membrane poration, whereas in the control simulation, the FTA aggregate quickly dissociated on the membrane surface, indicated by the drastic increase in $R_g$ (from 22 to 40 Å) and SASA (by 17%). Furthermore, the MG2 core readily intruded itself into the membrane, while TP1 stabilized the oligomer in the membrane (Figure 4 and S7). These events demonstrate how the arrangement of TP1 and MG2 within the aggregate structure can stabilize transmembrane pores. At the same time, inspection of Figures 4A3 and 4B3 established the formation of energetically stabilizing interactions between the positively charged sidechains of the AMP aggregates and the PG and CL headgroups. (2) The hetero-oligomers cause enhanced local disorder and facilitated membrane disruption. In the membrane-disrupted hetero-oligomer, MG2 readily adopted various orientations (Figure 4), which consequently altered the bound lipid orientations. Some lipid molecules can also be pulled out from the membrane by the MG2-TP1 oligomer (Figure 4A3).

It has been suggested that AMP oligomers at the nanometer scale provide a critical local concentration of peptides in functional conformations;[41] however, large AMP aggregates increase the free-energy cost for membrane poration and reduce the antibacterial activity.[42] In this respect, the electrostatic interactions between the MG2 and TP1 aggregates and the membranes observed in our simulations may provide stabilizing energy to enhance membrane disruption. Based on our computational and experimental results, we propose the following mechanism for the observed concentration dependence of the MG2-TP1 synergy: While the MG2-TP1 mixture at nanomolar to micromolar concentrations forms smaller, more active hetero-oligomers than individual peptides in solution, MG2 and TP1 — at a concentration higher than the naturally occurring ones — can generate larger hetero-oligomers and thus become less effective to embed into the membrane. Further, because MG2 aggregates more readily than TP1, excessive MG2 does not improve the antibacterial activity. Thus, the observation that the size of aggregates formed by TP1 and MG2 mixtures depends on concentration provides a pathway to regulate the synergetic behavior of the two AMPs. Moreover, this can be a general mechanism underlying the concentration dependence of synergistic AMPs, as aggregation/co-aggregation is ubiquitous among peptides. With awareness of such concentration dependences, we will likely find that synergistic AMPs exist more commonly than what is currently known.

In summary, by means of MD simulations and experimental characterizations, we explored the molecular origins of MG2-TP1 synergy. These two AMPs likely form oligomeric structures before binding to the anionic membrane surface. The hetero-oligomers enhance recognition to the bacterial membranes. Also, the hetero oligomers' high stability and significant membrane disruption potential helps stimulate water channel formation. Our studies provide valuable insight into the mechanistic details of AMP synergism, using the MG2-TP1 combination as a model system. Specifically, we show that co-aggregation plays an important role in enabling molecular interactions that may lead to the observed synergy. While only few synergistic AMPs are known currently, our findings from the MG2-TP1 combination suggest several possible directions to pursue for future synergistic AMPs, which can give rise to higher efficacy and lower risk of adverse effects.

**Supporting Information**. Simulation methods including model preparation, simulation protocols, and data analysis are included in the supporting information. Experimental methods describing peptide synthesis, dynamic light scattering, and the optical density measurements of E. coli inhibition are included as well.
The following files are available free of charge.
AMP_JPCL_SI.pdf file (PDF)

**Notes**

The authors declare no competing financial interests.

ACKNOWLEDGMENT

We thank PSC-ANTON, XSEDE, and the Vermont Advanced Computing Core for supercomputing resources. J. L., J. M. R., and C. L. were partially supported by an NIH award (R01GM129431). S.T.S. was supported by the U.S. Army Research Office (Grant 71015-CH-YIP).

Jacob M. Remington,[1] Chenyi Liao,[1] Mona Sharafi,[1] Emma Ste.Marie,[1,2] Jonathon B. Ferrell,[1] Robert Hondal,[1,2] Matthew J. Wargo,[3] Severin T. Schneebeli,[1] Jianing Li[1,*]

[1]Department of Chemistry, University of Vermont, Burlington, VT 05405

[2]Department of Biochemistry, University of Vermont, Burlington, VT 05405

[3]Department of Microbiology and Molecular Genetics, University of Vermont, Burlington, VT 05405

* Corresponding Author: Jianing Li (jianing.li@uvm.edu)

**Simulation Methods**

*Model Preparation.* The experimental structures from the Protein Data Bank (PDB) were used for the peptide (PDB IDs: 2MAG for MG2 and 2RTV for TP1). Major components of bacterial membranes, 1-Palmitoyl-2-Oleoyl-sn-glycero-3-PhosphoEthanolamine (PE), 1-Palmitoyl-2-Oleoyl-sn-glycero-3-PhosphoGlycerol (PG), and cardiolipin (CL), were used to build two bacterial membrane models: PE: PG (3:1), and PE:PG:CL (65:10:25, namely CL-rich) respectively (mass ratios in brackets). A typical model of the mammalian cell membrane was also used as an additional a control system, containing 1-Palmitoyl-2-Oleoyl-sn-glycero-3-PhosphoCholine (PC):cholesterol (89:11). The peptide models were prepared by Maestro (Schrödinger, *Inc*.), and combined with the membrane (on the XY plane) and solvent models by CHARMM-GUI.[1] The systems were neutralized by counterions in addition to 0.12 M NaCl. In our starting models, peptides were placed either in the solution arbitrarily (aggregation simulations) or above the upper leaflet of the lipid bilayer at a distance between 5 and 10 Å (permeation simulations). Summaries of all our constructs are provided in Tables S2-3. For steering models, larger hetero-oligomers (~70 Å in diameter) – formed by combining four MG2-TP1 aggregates obtained by ANTON simulations – were used to closely match the hydrodynamic radius observed in the DLS measurements (Fig. S1).

*Simulation Protocols.* Our all-atom simulations were performed with the CHARMM36 force field,[2] <u>which has been extensively tested for protein-membrane systems, implemented in Gromacs[3] or Amber[4]</u> with the TIP3P water model. However, the aggregation of MG2 and TP1 were simulated with classical MD by the specialized Anton supercomputer. Both equilibration and production runs were carried out in the NPT ensemble (298K, 1 bar, Langevin thermostat and Nose-Hoover Langevin barostat) with a time step of 2 fs, while the steered MD simulations were in the NVT ensemble. Isotropic and semi-isotropic barostats were used for aggregation and permeation simulations respectively. The particle mesh Ewald (PME) technique was used for the electrostatic calculations. The van der Waals and short-range electrostatics were cut off at 12.0 Å with the switch at 10.0 Å.

Peptide binding to the membrane surface was modeled with 100- to 200-ns classical MD simulations, followed by the permeation simulations used a new protocol that is comprised of alternating steered and classical MD steps. This protocol was designed to mimic the process of peptides colliding into the elastic lipid bilayer. In each step, we first applied constant velocity pulling on the backbone atoms of the peptides along the negative Z-axis direction (toward the lower leaflet) for 40 ps to mimic the slow collision, followed by an 8-ns classical MD simulation to relax the peptide-lipid interactions. A Fortran script was compiled for implementing this collective variable in Amber and was found to reproduce the SMD induced pore formation for the same system implemented in Gromacs. Typically, a steered process for 30 to 40 ps towards the membrane surface generates approximately 4.5-nm relative displacement of peptides towards the membrane surface with a soft membrane curvature (Fig. S4). For the optimal setting of the steered MD simulations, several pulling trials were carried out. We ultimately chose a force constant of 20 kcal/mol·Å$^2$ and a pulling rate of 0.005 Å/timestep (2.5 Å/ps), since this setting provides a strong enough collision to attack the membrane integrity. In total, we carried out near 20 steps for each simulation, during which the thickness of the hydrophobic region beneath the peptides was reduced (which we consider as an indicator of lowered barrier for permeation). When we observed initial membrane permeation after a steered and classical MD step, an extended classical MD simulation step was applied for 150 ns to ensure that the pore is stable.

*Data Analysis.* Analysis of all the simulations was carried out with in-house TCL scripts in VMD and Python tools, while model visualization was performed with VMD and Pymol (Schrödinger, Inc.). Polar contacts within 3.6 Å were shown by Pymol (Schrödinger, Inc.). The onset of the pore state was indicated, when a peptide from the oligomer is within 10 Å from the bottom of the lower leaflet and water molecules are present in the membrane hydrophobic core. The membrane hydrophobic thickness was obtained by dividing the membrane into 4×4 units on the XY-plane; in each unit we calculated the average distance between the three close water molecules above the center-of-mass (COM) of the upper lipid tails and another three close water molecules below the COM of the lower lipid tails. Radius of Gyration and solvent accessible surface area were calculated using either VMD or cpptraj for the Gromacs/Anton and Amber simulations respectively. The number of assembled peptides was calculated by finding the largest number of peptides self-associated (i.e. with any atomic distance < 7 Å) implemented in VMD. Helicity and random coil were also calculated using VMD scripts.

**Experimental Methods**

*Peptide Synthesis.* MG2 was purchased from Gen Script (> 95% purity). TP1 was synthesized using Fmoc-based solid phase peptide synthesis (SPPS)[5,6] and orthogonal cysteine (Cys) protection[7-9]. Notably, the carboxyl-terminal end of Tachyplesin contains an arginine α-amide[10]. To achieve this C-term amide functionality, Rink Amide AM resin (100-200 mesh, Novabiochem®) was used for SPPS. Tachyplesin also contains two disulfide bonds: one between Cys3 & Cys16, and one between Cys7 & Cys12[10]. To ensure correct disulfide connectivity, which can often be a synthetic challenge in peptides containing multiple disulfide bonds[7-9], methods developed by Ste.Marie and coworkers were utilized[11]. The first disulfide bond was formed between Cys7 & Cys12, which were *S*-trityl (*S*-Trt) protected during SPPS. Upon cleavage from the resin using 96:2:2 (trifluoroacetic acid: triisopropylsilane: H$_2$O), the Trt protecting groups were removed via acidolysis. The resulting Cys-thiols were air oxidized in ammonium bicarbonate buffer (pH 8.12), forming the first disulfide bond. The peptide was then frozen and lyophilized to remove water and volatile buffer salts. Next, the second disulfide bond was formed between Cys3 & Cys16, which were *S*-acetamidomethyl (*S*-Acm) protected during SPPS. Lyophilized peptide powder was re-dissolved in 96:2:2 (trifluoroacetic acid:triisopropylsilane:H$_2$O) containing 5-fold excess 2,2'-dipyridyl diselenide (PySeSePy), and allowed to react for 8 h at 37 °C. The addition of PySeSePy to cleavage cocktail was previously shown to remove *S*-Acm protecting groups with concomitant disulfide bond formation during the synthesis of guanylin[11], and we found that this method worked well for the installment of the second disulfide bond of Tachyplesin. After incubation with PySeSePy, the peptide was precipitated with cold ether and lyophilized to powder. Excess PySeSePy was subsequently removed via preparative high-pressure liquid chromatography (HPLC) purification using a water/acetonitrile gradient. The correct mass was confirmed for each step of the synthesis using MS analysis, and the spectra are given as Fig. S8-S10.

*Dynamic light scattering (DLS).* DLS experiments were performed at room temperature with Dynamic light scattering (DLS) measurements were acquired on a Malvern Panalytical Zetasizer Nano ZSP instrument, using a Hellma quartz cuvette (ZEN2112). The raw DLS data was processed and analyzed with the Zetasizer software (version 7.02). For all the measurements, solutions of $2.0 \times 10^{-7}$ to $10^{-11}$ M MG2:TP1 in HEPES buffer at pH = 7.0 ± 0.1 have been prepared in advance and used promptly. All volumetric measurements for solution preparation were performed with Rainin Positive Displacement (MR-10, -100, -1000) micropipettes. To investigate the size of the aggregates, first the DLS of TP1 has been recorded. This was followed by the addition of MG2 in the same concentration as TP1 to the solution. After waiting for about 60 seconds, the DLS of the resulting mixture was reported in 3 minutes scanning time. Finally, the spectrum of individual MG2 has been obtained and the resulting hydrodynamic diameter has been compared with the corresponding ones in mixture and TP1.

*Optical density (OD) measurements of E. coli inhibition.* *Escherichia coli* strains were maintained on LB medium. Growth of all species was measured based on the optical density at 600 nm ($OD_{600}$) after growth for 24 h in MOPS minimal medium with a 20 mM concentration of the sole carbon source, as described previously.[12] all cultures were started at an OD600 of 0.01. All the experiments have three replicas. For every pair of data points in Figure 2B, statistical significance of a difference in measured mean was tested using a t-test with the Holm-Sidak method, $\alpha = 0.05$, and the assumption that all trials were sampled from populations with the same scatter.

**Table S1.** Relative *E. coli* growth* under combinations of TP1 and MG2 AMPs**, trial 1.

| % | 0 µg/ml TP1 | 2 µg/ml TP1 | 8 µg/ml TP1 | 32 µg/ml TP1 |
|---|---|---|---|---|
| 0 µg/ml MG2 | 100 | 95.24 | 85.08 | 65.08 |
| 2 µg/ml MG2 | 93.70 | 90.51 (89.23) | 78.04 (79.72) | 52.02 (60.98) |
| 8 µg/ml MG2 | 88.98 | 81.32 (84.74) | 69.27 (75.70) | 53.94 (57.91) |
| 32 µg/ml MG2 | 87.80 | 83.65 (83.62) | 76.72 (74.70) | 55.67 (57.14) |

*Relative growth was calculated by optimal density (OD) at 600 nm.
**Parenthesis represent predicted synergy using Equation 1.

**Table S2.** Relative *E. coli* growth* under combinations of TP1 and MG2 AMPs**, trial 2.

| % | 0 µg/ml TP1 | 2 µg/ml TP1 | 8 µg/ml TP1 | 32 µg/ml TP1 |
|---|---|---|---|---|
| 0 µg/ml MG2 | 100 | 96.79 | 89.38 | 66.17 |
| 2 µg/ml MG2 | 100.70 | 94.01 (97.57) | 83.13 (90.00) | 64.31 (66.64) |
| 8 µg/ml MG2 | 89.13 | 84.75 (86.27) | 67.18 (79.67) | 52.23 (58.98) |
| 32 µg/ml MG2 | 91.11 | 85.70 (88.18) | 78.53 (81.44) | 58.51 (60.29) |

*Relative growth was calculated by optimal density (OD) at 600 nm.
**Parenthesis represent predicted synergy using Equation 1.

**Table S3.** Relative *E. coli* growth* under combinations of TP1 and MG2 AMPs**, trial 3.

| % | 0 µg/ml TP1 | 2 µg/ml TP1 | 8 µg/ml TP1 | 32 µg/ml TP1 |
|---|---|---|---|---|

| 0 μg/ml MG2 | 100 | 99.89 | 90.56 | 71.29 |
| 2 μg/ml MG2 | 99.53 | 88.12 (99.42) | 72.18 (90.13) | 61.80 (70.95) |
| 8 μg/ml MG2 | 93.61 | 72.62 (93.51) | 64.91 (84.77) | 51.19 (66.73) |
| 32 μg/ml MG2 | 92.75 | 74.45 (92.65) | 68.96 (83.99) | 57.85 (66.12) |

*Relative growth was calculated by optimal density (OD) at 600 nm.

**Parenthesis represent predicted synergy using Equation 1.

**Table S4.** Simulation summary of MG2 and TP1 aggregation in the solution and on the membrane surface. Each unbiased simulation was carried out until the size of assemblies did not change. While MG2 and TP1 formed hetero-oligomers in all solution simulations (Y1-Y3), they did not show consistent aggregation on the membrane surface (B7 & B9).

| System | Simulation Components | box size (Å$^3$) | No. of atoms | Simulation time per replica (ns) |
|---|---|---|---|---|
| Y1 | 4 random-coiled MG2 + 4 cyclic TP1 | 85×85×85 | 60391 | 2000 (×2) |
| Y2 | 4 helical MG2 + 4 linear TP1 | 85×85×85 | 57074 | 200(×2) |
| Y3 | 4 helical MG2 + 4 cyclic TP1 | 60×60×60 | 19134 | 800 (×2) |
| B7 | PE: PG 4:1, dispersed 4 MG2 + 4 TP1 | 91×91×88 | 74878 | 1976 |
| B9 | PE: PG 4:1, complex 4 MG2 + 1 TP1 | 62×63×99 | 35404 | 1169 |
| B11 | 16 FTA | 67×67×67 | 31451 | 1000 |

**Table S5.** Simulation summary of MG2 and TP1 for SMD induced membrane poration. The starting peptide oligomers were extracted from the final snapshot of the corresponding solution simulations. Membrane pores formed in the simulations with bacterial membrane models (B2 & B3), but was not observed in the control simulations with non-AMP peptide FTA (B10) or with a mammalian membrane model (M1).

| System | Simulation Components | box size (Å$^3$) | No. of atoms | Simulation time per replica (ns) |
|---|---|---|---|---|
| B2 | PE:PG 3:1, 12 MG2 + 4 TP1 | 86×86×94 | 72994 | 800 (×2) |
| B3 | PE:PG:CL 65:10:25 12 MG2 + 4 TP1 | 89×89×88 | 72415 | 600(×2) |
| B12 | PE:PG 3:1, 16 FTA | 101×101×108 | 113132 | 600(×2) |
| M1 | PC:Cholesterol 89:11, 12 MG2 + 4 TP1 | 103×103×108 | 119704 | 500(x2) |

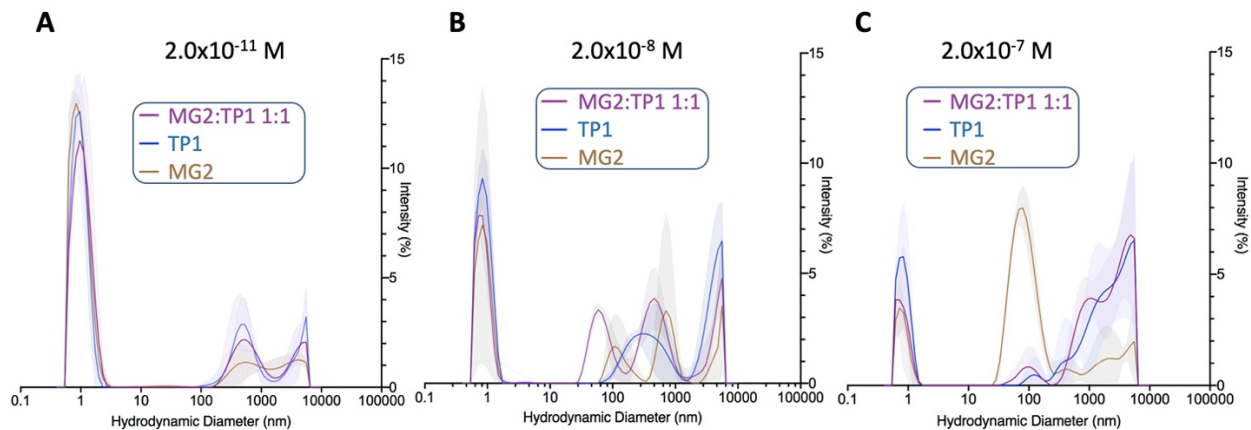

**Figure S1.** DLS measurements at three total peptide concentrations.

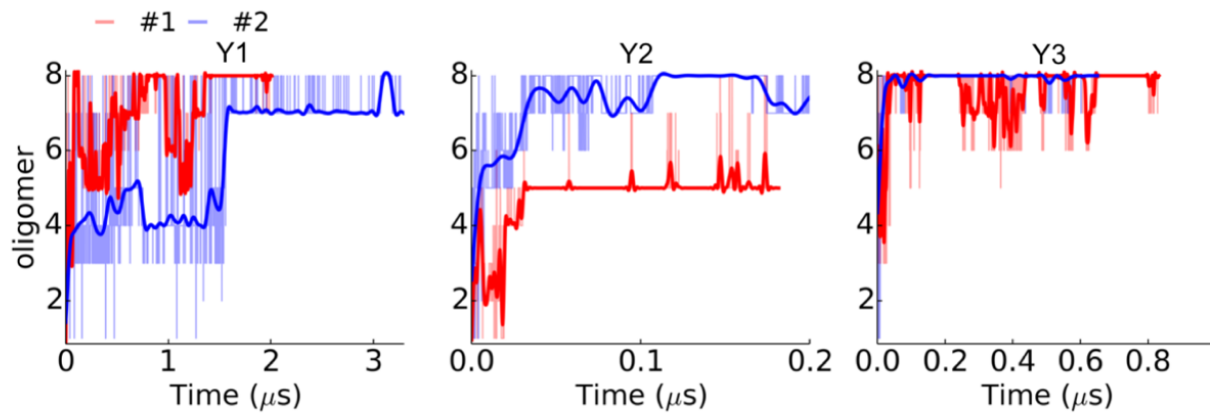

**Figure S2.** The number of assembled peptides in time series for systems Y1, Y2, and Y3, each of which displays two replicas.

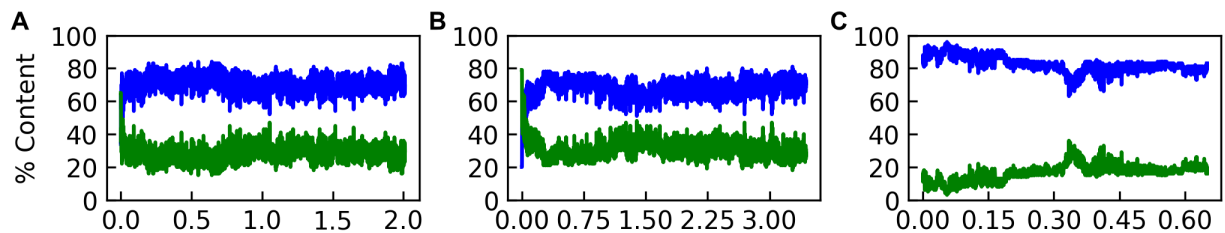

**Figure S3.** The percent content of helicity (blue) and coil (green) of MG2 (A-C) are shown during the aggregation of systems S1, S2, and S3 in panels A-C respectively.

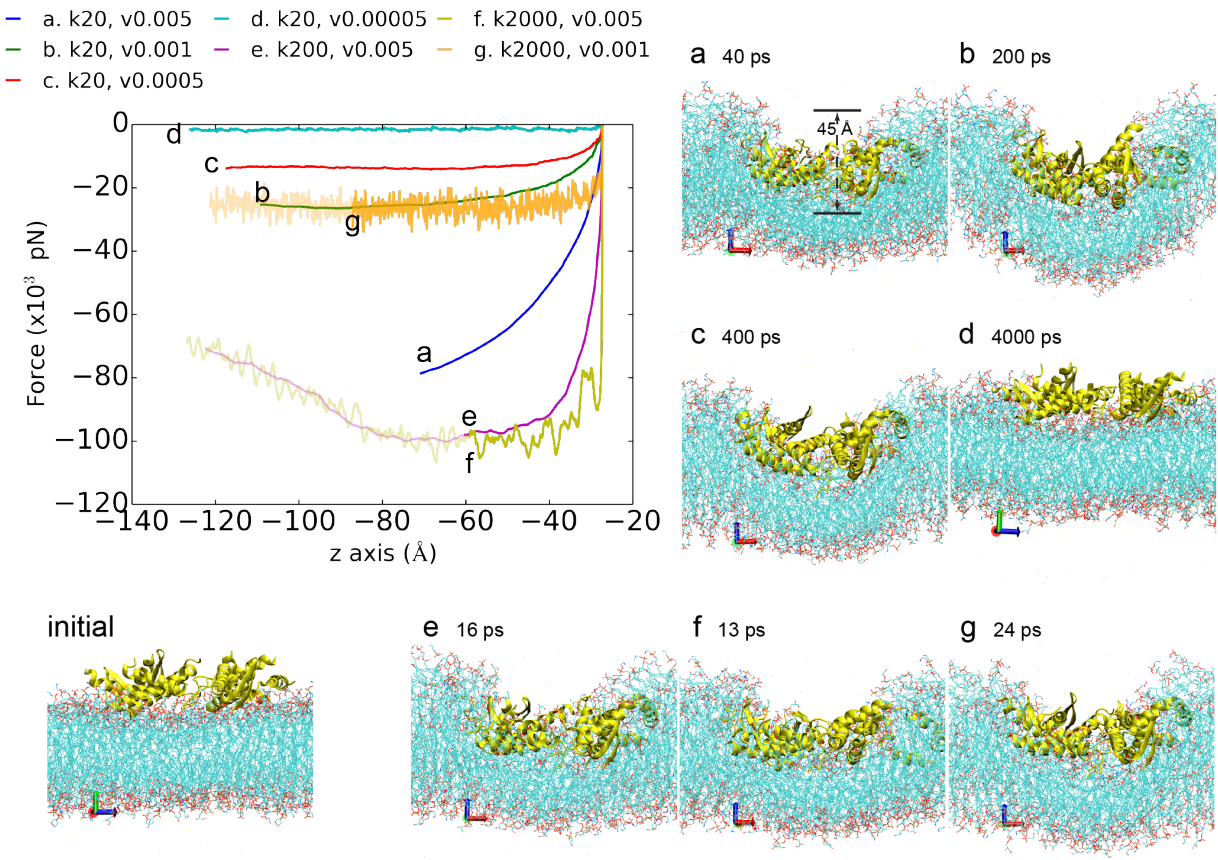

**Figure S4.** Force versus position (the peptides center-of-mass position along the negative z axis) plots using different spring constant ($k$, kcal/mol·Å$^2$) and pulling velocities ($v$, Å/timestep) in the mixed peptides-membrane system. Snapshots in *a-g* pulling trials were shown on the right and bottom and labeled on the curves where they appeared. The most appropriate value for spring constant is 20 kcal/mol·Å$^2$, because spring constant of 2000 kcal/mol·Å$^2$ induce large bias (noise/fluctuation shown in the curves *f* and *g*) to the system. Continuing pulling in *e*, *f*, and *g* eventually leads to breaking through of the membrane, which is depicted in the half transparent color of the traces. When the pulling rate was slowed down 100 times, from 0.005 to 0.00005 Å/timestep (*d*), the pulling approaches a process during which translocation of the entire peptides-membrane complex along the negative z axis was observed with mild relative movement of peptides towards the membrane. Considering the pulling rate did not affect the stability of the system, but will increase the overall translocation along z axis, we chose a force constant of 20 kcal/mol·Å$^2$ and a pulling rate of 0.005 Å/timestep to accomplish a 40 ps steering process. The referring state (*a*) was chosen for two general reasons: (i) the center-of-mass of mixed peptides is roughly close to but not beyond the middle lipid bilayer; (ii) the slope of the force versus position curve is leveling out.

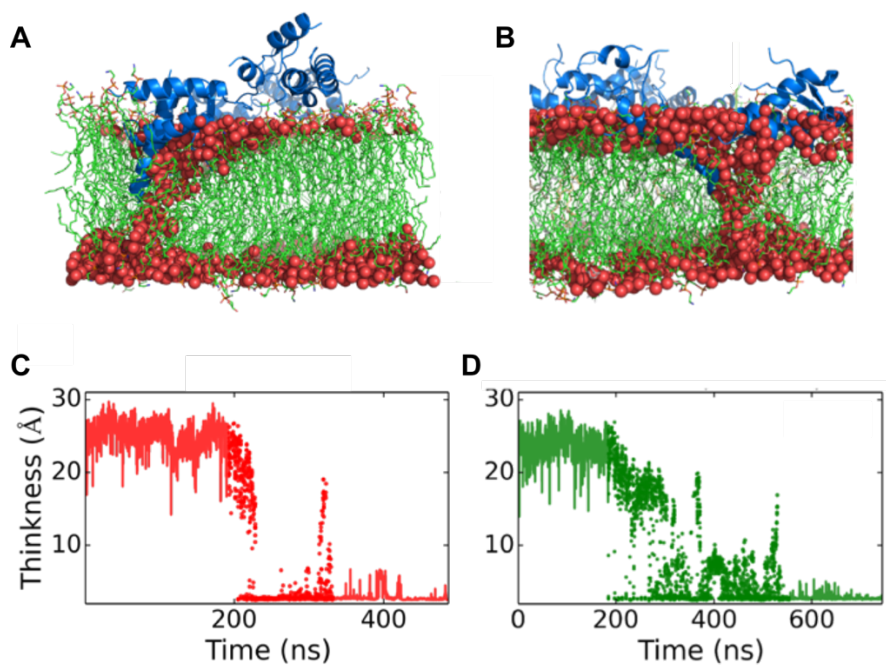

**Figure S5.** Final snapshots from replica simulations of pores formed in the POPE:POPG 3:1 membrane **(A)** and the cardiolipin rich membrane **(B)** after simulation of interactions with the MG2-TP1 hetero-oligomer. Time traces of the membrane thickness for these respective systems are shown below each with the normal MD (solid line in C and D) and alternating SMD and MD (dots in C and D) highlighted. Red spheres represent the oxygen atoms from the water molecules near or within the membrane.

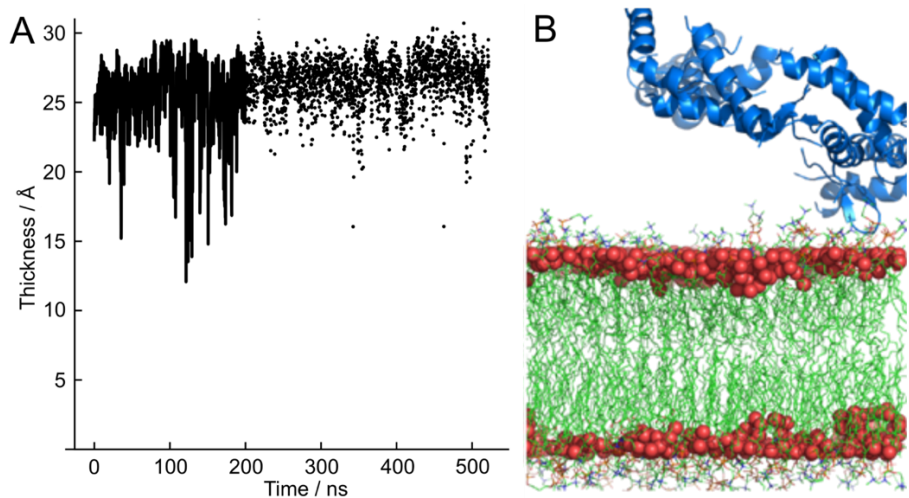

**Figure S6.** Final snapshot of a simulation of the MG2-TP1 hetero-oligomer (blue cartoon) with a mammalian membrane model (green lines) containing POPC:Cholesterol at the 89:11 mass ratio. Red spheres represent the oxygen atoms from the water molecules near or within the membrane.

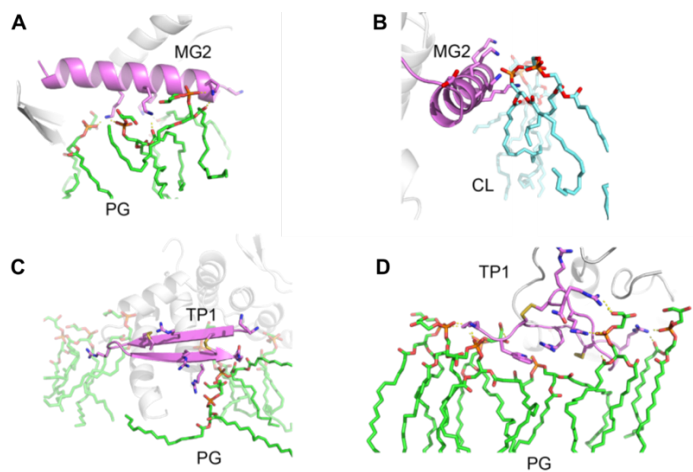

**Figure S7.** Strong anionic lipid-residue interactions between **(A)** MG2 and POPG, **(B)** MG2 and cardiolipin (CL), **(C)** TP1 and POPG, and **(D)** TP1 and POPG, which were identified from the final snapshots.

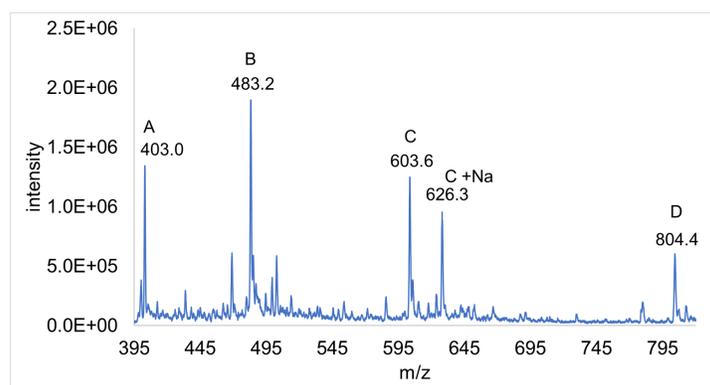

**Figure S8.** Tachyplesin peptide after cleavage from the resin. Cys7 and Cys12 are in Cys-thiol form, while Cys3 and Cys16 are S-Acm protected. Due to the high Arg and Lys content of this peptide, which pick up positive charges readily, the peptide is detected as a mixture of multiply charged species. Species "A" is the $m/z^{+6}$ product (theoretical $m/z^{+6}$ = 402.76); species "B" is the $m/z^{+5}$ product (theoretical $m/z^{+5}$ = 482.63); species "C" is the $m/z^{+4}$ product (theoretical $m/z^{+4}$ = 603.04); species "C +Na" is the sodium adduct of the $m/z^{+4}$ product (theoretical $m/z^{+4}$ + $Na^+$ = 626.60); and species "D" is the $m/z^{+3}$ product (theoretical $m/z^{+3}$ = 803.92). The major peaks are labelled in the spectrum with the observed m/z value.

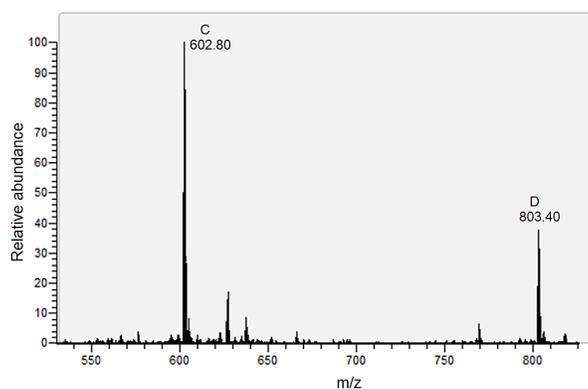

**Figure S9.** Tachyplesin peptide with first disulfide bond formed. Tachyplesin after formation of first disulfide bond between Cys7 & Cys12 by air oxidation. Cys3 & Cys16 remain *S*-Acm protected. Species "C" is the $m/z^{+4}$ product (theoretical $m/z^{+4}$ = 602.54); and species "D" is the $m/z^{+3}$ product (theoretical $m/z^{+3}$ = 803.06). The major peaks are labelled in the spectrum with the observed m/z value.

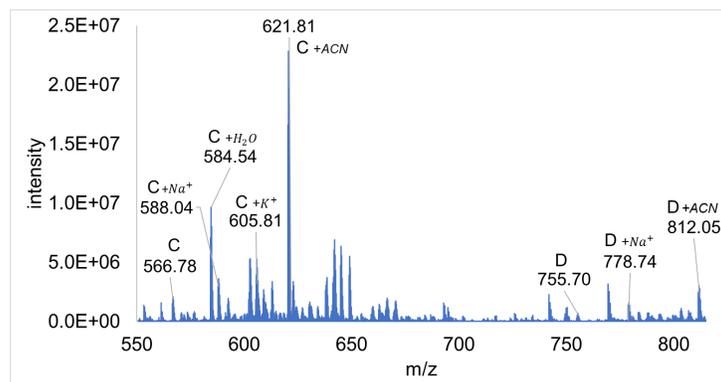

**Figure S10.** Tachyplesin peptide with both disulfide bonds formed. Tachyplesin after formation of second disulfide bond between Cys3 & Cys16 using PySeSePy, and after HPLC purification using a gradient of water/acetonitrile (ACN). Species "C" is the m/z$^{+4}$ product (theoretical m/z$^{+4}$ = 566.50); species "C +Na$^+$" is the sodium adduct of the m/z$^{+4}$ product (theoretical m/z$^{+4}$ + Na$^+$ = 588.5); species "C +H$_2$O" is the monohydrate of the m/z$^{+4}$ product (theoretical m/z$^{+4}$ + H$_2$O = 584.50); species "C +K$^+$" is the potassium adduct of the m/z$^{+4}$ product (theoretical m/z$^{+4}$ + K$^+$ = 505.60); species "C +*ACN*" is the acetonitrile/NH$_3^+$ adduct of the m/z$^{+4}$ product (theoretical m/z$^{+4}$ + ACN/NH$_3^+$ = 622.6); species "D" is the m/z$^{+3}$ product (theoretical m/z$^{+3}$ = 755.00); species "D +Na$^+$" is the sodium adduct of the m/z$^{+3}$ product (theoretical m/z$^{+3}$ + Na$^+$ = 778.00); and species "D +*ACN*" is the acetonitrile/NH$_3^+$ adduct of the m/z$^{+3}$ product (theoretical m/z$^{+3}$ + ACN/NH$_3^+$ = 811.10). Note: the ACN adduct is prevalent in our final product due to the HPLC purification step where ACN is the solvent. The major peaks are labelled in the spectrum with the observed m/z value.